\documentclass[12pt]{article}

\usepackage{amsmath,amssymb,amsfonts,graphicx}
\usepackage{epsfig}

\newcommand {\beq} {\begin{equation}}
\newcommand {\eeq} {\end{equation}}
\newcommand {\beqn}{\begin{eqnarray}}
\newcommand {\eeqn} {\end{eqnarray}}
\newcommand{\bea}{\begin{eqnarray}}
\newcommand{\eea}{\end{eqnarray}}

\def\ie{\emph{ie.}~}

\def\1{\mathbbm{1}}
\def\N{{\cal N}}

\def\T{{\cal T}}
\def\tL{\widetilde{L}}
\def\tz{\widetilde{z}}

\newcommand\blfootnote[1]{%
  \begingroup
  \renewcommand\thefootnote{}\footnote{#1}%
  \addtocounter{footnote}{-1}%
  \endgroup
}

\begin{document}

\begin{titlepage}

\flushright{Imperial/TP/2012/ER/01}

\vskip 2mm

\begin{center}
{\LARGE  Comments on Critical Electric and Magnetic Fields from Holography  } %
\blfootnote{\it  E-mail: {stefanobolo@gmail.com, kiefer@iap.fr, eliezer@vms.huji.ac.il}}

\vskip 8mm
{\large S. Bolognesi${}^a$, F. Kiefer${}^{a,b}$ and E. Rabinovici${}^{a,c}$ }

\vskip 2mm
{\small \it

${}^a$Racah Institute of Physics, The Hebrew University of Jerusalem, 91904, Israel

${}^b$GRECO, Institut d'Astrophysique de Paris
98bis Bd Arago, 75014 Paris, France

${}^c$Theoretical Physics Group, The Blackett Laboratory, Imperial College London,
Prince Consort Road, London SW7 2AZ, UK

}

\end{center}

\begin{abstract}

We discuss some aspects of critical electric and magnetic fields in a field theory with  holographic dual description. We extend the analysis of \cite{Semenoff:2011ng}, which finds a critical electric field at which the Schwinger pair production barrier drops to zero, to the case of magnetic fields. We first find that, unlike ordinary weakly coupled theories, the  magnetic field is not subject to any perturbative instability  originating from the presence of a tachyonic ground state in the W-boson spectrum.  This follows from the large value of the 't Hooft coupling $\lambda$, which prevents the Zeeman interaction term to overcome the particle mass at high $B$.  Consequently, we study the next possible B-field instability, i.e.  monopole pair production, which is the S-dual version of the Schwinger effect. Also in this case a critical magnetic field is expected when the tunneling barrier drops to zero. These Schwinger-type criticalities are the holographic duals, in the bulk, to  the fields $E$ or $B$  reaching the tension of  F1 or D1 strings respectively.  We then discuss how this effect is modified  when electric and magnetic fields  are present simultaneously and dyonic states in the spectrum can be pair produced by a generic $E - B$ background. Finally, we analyze finite temperature effects on Schwinger criticalities,  i.e.  in the AdS-Schwarzshild black hole background.  

\end{abstract}

\vfill
\end{titlepage}

\numberwithin{equation}{section}
\tableofcontents
\renewcommand{\thefootnote}{\#\arabic{footnote}}
\setcounter{footnote}{0}

\setcounter{equation}{0}

\section{Introduction}

Low energy limits of string theories were originally found to be very similar 
to Quantum Field Theories. Later some QFTs were found to be secretly
equivalent to some string theories. There are many subtle and less subtle
differences between string theories and field theories in flat space time. One
such difference emerges in the presence of electric and magnetic backgrounds.

In QED on one hand the presence of an electric field leads  non perturbatively
to electron-positron pair production, yet this  does not seen to produce
a threat to the vacuum stability, the electric field can be increased at will. The application of a magnetic field on a QED like system containing electrically charged  spin zero or spin one-half particles has neither perturbative nor non perturbative destabilizing effects. If
monopoles happen to exist in the field theory, they will be pair produced in the magnetic field
yet this process will not render the vacuum unstable and will not set a limit on the possible value of the
magnetic field. The presence of a spin one charged particle, such as the W bosons, will instead lead to a perturbative instability.

 In string theory on the other hand the extended object properties of the states
results in an appearance of a maximally allowed value for the electric field.
In the presence of the magnetic field in flat space the roles are reversed with respect to the field theories.  The perturbative magnetic instability due to the presence of higher spin particles is much softened by the extended nature of strings, and in many cases can also  completely disappear in some circumstances.

As one moves from flat space to an AdS like space which has a holographic
field theory dual, such a discrepancy can no longer be tolerated in some regions
of the parameter space. The question becomes if  it is the field theory or
the string theory character which will determine the behavior of the system
in the presence of the electromagnetic fields. It was found in \cite{Semenoff:2011ng}  that it is
the stringy character which dominates in the presence of the electric field.
In this work we study the  issue in the presence of a magnetic field with and without the electric field.

A common phenomenon in quantum field theories is the Schwinger effect: the pair production of charged pairs, particle $q$ and anti-particle $\bar{q}$, induced by a background electric field. The production rate probability per unit of time  at leading order is
\beq
\label{pp}
w \propto e^{-\pi m^2 / e E}
\eeq
where $m$ is the particle mass, $e$ the charge and $E$ the electric field. The electric field constantly  pulls out of the vacuum  $q \ \bar{q}$  pairs with a well defined probability. Even when the exponent $\pi m^2/e E$ becomes of order one and thus the production rate becomes large, there are no indications of  phase transitions. Just the formula (\ref{pp}) needs  to be corrected with higher order terms corresponding to multiple pair production.

In the Schwinger effect we see  a clear-cut distinction between QED-like field  theories and string theories.  In the latter,  with electric charges that sit  at the  extremities of the open string,  there is a critical electric field at which a phase transition occurs.  This is when the force  applied by the electric field on either of the charged ends becomes equal to the string tension and  the string thus breaks apart \cite{Fradkin:1985qd,Burgess:1986dw,Bachas:1992bh}:
\beq
\label{Ecriticalstring}
e E_{cr} = \frac{1}{2 \pi \alpha'} \ .
\eeq
At this value of $E$  the pair production barrier drops to zero we cannot keep increasing the electric field  any longer. This is a real break down,  the effective tension becomes zero and the usual string description inappropriate. Note moreover that this criticality happens also for neutral strings with opposite charges at the extremities and no Schwinger pair creation \cite{Burgess:1986dw,Bachas:1992bh,Ambjorn:2000yr}.

But we know that some  field theories are  also equivalent to string theories. In this case we would expect the field theory to behave in a ``stringy way'' when tested with increasingly high electric fields, and thus to  have  a criticality at a certain finite value $E_{cr}$, similar to (\ref{Ecriticalstring}), although now expressed exclusively in term of field theory parameters.  This problem was analyzed in \cite{Semenoff:2011ng,Gorsky:2001up}. They considered $\N=4$ super Yang-Mills in the Coulomb branch  with a partial symmetry breaking $SU(N+1) \to SU(N) \times U(1)$,  where the $U(1)$ massless gauge boson is to be thought as the above electro-magnetic field and the massive $W$ bosons plays the role of the charged particle to be pair created by the electric field.  When the gravity dual description is weakly coupled, that is large $N$ and large 't Hooft coupling, they showed that the Schwinger pair production develops indeed a critical instability at 
\beq
\label{Ecr}
E_{cr} = \frac{2 \pi m^2}{\sqrt{\lambda}} \ ,
\eeq
where $\lambda$ is the 't Hooft coupling and $m$ the $W$ boson mass \cite{Semenoff:2011ng}. This is the holographic dual to  the fact that in the bulk of AdS strings reach a critical point when the electric field breaks them apart (\ref{Ecriticalstring}).

We want here to analyze different aspects of this problem. The first generalization one could ask is what happens if the system is probed with an external magnetic field instead of an electric field. A magnetic field is known to  induce a dual version of the Schwinger effect if the theory admits magnetic charged particles, such as 't Hooft-Polyakov monopoles. This was studied in \cite{Affleck:1981ag,Affleck:1981bma} by computing the Euclidean worldline path integral of monopole loops. The computational method turned out to be applicable also to the electric Schwinger effect  with S duality. Let's take the $SU(2)$ gauge theory with adjoint scalar field and Lagrangian $L=\frac{1}{4} F^2 + (D\phi)^2 -V(\phi)$ and $D = \partial -ig A$.
 In the Higgs phase for the adjoint scalar field, where $\langle \phi \rangle = v$ and  the gauge group is broken to  $U(1)$, we have in the spectrum the  $W$ boson with mass $m_W = g v$ and the monopole with mass $m_{M} = 4 \pi v /g $.  Electric field induces Schwinger production of $W^+ \  W^-$ pairs at a rate 
\beq
\label{pairpw}
w_{E \to W\bar{W}} \propto e^{-\pi m_W^2/ g E} \ .
\eeq
Similarly a magnetic field induces monopole-antimonopole pair production at a rate 
\beq
\label{pairpm}
w_{B\to M\bar{M}} \propto e^{- \pi  m_M^2 / \widetilde{g} B} \ ,
\eeq
where $\widetilde{g} = 4 \pi / g$ is the magnetic coupling. The two processes are S-dual to each other, with the S-duality transformation given by $E \to B$, $v \to v$, and $ g \to \widetilde{g}$. We thus have two typical scales, which we call $E_{Schw}$ and $B_{Schw}$, where the Schwinger pair productions become strong:
\beq
\label{schscales}
E_{Schw} \simeq \frac{\pi m_{W}^2}{g} = \pi g v^2 \ , \qquad \quad B_{Schw} \simeq \frac{m_{M}^2 g}{4} = \frac{4 \pi^2 v^2}{ g} \ .
\eeq
Here by strong we mean that the probability is no longer exponentially suppressed  and beyond these values the semiclassical treatment is no longer valid.\footnote{In QED $E_{Schw} \simeq 10^6 E_{Ion}$ where $E_{Ion}$ is the typical electric field that is necessary to ionize an atom. This is why the Schwinger effect has not yet been experimentally observed, although this may soon change. See for example \cite{Dunne:2008kc} for a review.}   Note that assuming weak coupling $g \ll 1$  and  comparable  magnitudes of field strengths  $E \simeq B$,  the monopole pair production is suppressed with respect to the $W$ pair production. The are two competing effects here. The monopole coupling is higher by a factor $1/g^2$, but the monopole mass squared is also higher by a factor $1/g^4$ and this dominates.  The two scales (\ref{schscales}) are then related by the ratio $E_{Schw}: B_{Schw} = 1:1/g^2$.

For the magnetic field there is also another effect which should be taken into account, and turns out to be much more important than the monopole pair production. This is the gyromagnetic instability of the $W$ bosons spectrum. 
Fermions or scalar fields have no instability in the magnetic field background. In QED for example nothing prevents from increasing $B$ to whatever value. this is not true instead for spin $1$ W bosons. This instability is due to the Zeeman effect, the coupling between spin and magnetic field proportional  to the gyromagnetic factor $g_S$.  For the $W$ boson this factor is $ g_S= 2$, exactly the same as the fermion, although its spin is twice.  The energy squared of the $W$ bosons is given by the classical solution of the Proca wave equation in the $B$ field background. It is  given by the Landau level term plus the Zeeman term  plus the bare mass term: 
\beq
\label{spectrumfield}
{\cal E}_{n, \uparrow \downarrow}^2 = (2n +1) g B  \pm  g_S g B \cdot S + m_{W}^2 \ .
\eeq
The ground state is $n=0$ with the spin-down along $B$ and  its square energy is ${\cal E}_{0, \downarrow}^2 = -g B +m_{W}^2$. The Zeeman splitting between spin-up and spin-down states is $2g_S g B S$. Thus there is always a critical magnetic field, which we denote  $B_{gyro}$\footnote{For the lightest charged spin-1 particle, the $\rho$ meson of QCD, the critical field is  $B_{gyro} \simeq 10^{16}$ Tesla and thus very high indeed, although it can be reached in heavy ion collision (see for example \cite{Chernodub:2010qx,Bali:2012zg}).  We thank F.~Bruckmann and Z.~Komargodski for discussions about this issue.}, above which the ground state is tachyonic
\beq
\label{Bgyro}
B_{gyro} = \frac{m_{W}^2}{g} = g v^2 \ .
\eeq  
The existence of this instability is precisely due to the fact that the gyromagnetic factor is $g_S=2$ and thus bigger than $1/S$. For an electron there is no such an instability.  
This is a vacuum instability, the ground state becoming tachyonic is the signal of a phase transition which can be driven by the $W$ condensate \cite{Ambjorn:1988gb}\footnote{The other  phase  is in general believed to be the unbroken phase $\langle\phi\rangle =0$ where the non-Abelian $SU(2)$ is restored and the transition happens through vortex formations driven by the $W$ condensate.}.
Note that $B_{gyro} $ is much smaller than $B_{Schw}$ and  thus, as we increase $B$, the gyromagnetic instability is reached long before the monopole antimonopole pair production has any chance to  become strong. So the  story for weakly coupled theories  is  the following. Schwinger pair production is the dominant effect at large $E$ and becomes strong at $E_{Schw}$. For the magnetic field instead the monopole-antimonopole pair production is not the dominant effect. The $B_{gyro}$ instability is the most important effect, it is the one encountered first.

Let us draw a comment about the perturbative versus non-perturbative nature of those effects. The gyromagnetic instability (\ref{Bgyro}) is a perturbative effect, it is the emergence of a tachyonic state in  the spectrum of W-bosons (\ref{spectrumfield}) and can be seen in perturbative expansion. The Schwinger effect of W-bosons due to the electric field is instead non-perturbative and can not be seen in any perturbative expansion being exponentially suppressed like $e^{-1/g}$.

The main part of the paper is devoted to the analysis of the same phenomenon when the field theory is described by a holographic string theory in AdS space.  One surprise  found   in \cite{Semenoff:2011ng}  is the emergence of a  critical $E_{cr}$ due to the string breakdown.   We find another surprise for the magnetic side of the problem.  First there is no longer a gyromagnetic instability, this  effect simply disappears.  The stringy behavior is again responsible for this disappearance, in a similar way it was responsible for the existence of $E_{cr}$, through the $\alpha'$ corrections.   The formula (\ref{spectrumfield}) for the W boson spectrum is valid only in the field theory limit and in general it receives large  $\alpha'$  corrections as the dimensionless coupling  $g B \alpha'$ becomes big  \cite{Abouelsaood:1986gd,Ferrara:1993sq}. The field theory instability in the presence of a magnetic field remains in string theory in a flat space time background as such a theory contains Regge trajectories of  particles with higher spin.\footnote{Magnetic instabilities in AdS are also considered in different set-ups in which the fields lives in the bulk and not on a  brane, see for example \cite{Ammon:2011je,Almuhairi:2011ws,Donos:2011pn}. In those cases $\alpha'$ corrections are not relevant. }  We will explain in the paper why those corrections  completely wipe out the  tachyonic instability of the ground state, even if the  magnetic field is taken arbitrarily large.

The absence of the gyromagnetic instability opens the possibility for the monopole pair production to become significant. The rate of monopole pair production can be computed in a similar way to the W boson pair production and a critical magnetic field $B_{cr}$ exists  where the monopole-antimonopole barrier drops to zero.  This is the S-dual version of the previous effect. It happens when the $B$ field reaches the magnitude needed to break the D1-string. Although the D1 is much heavier than the F1, the coupling of the B field is also stronger with the same factor, thus $B_{cr} =  E_{cr}$. This phenomenon was first discussed for flat space-time in \cite{Gorsky:2001up}.  We then also discuss the pair production of dyons and the effect of having a generic $E$ and $B$ background simultaneously present.  We will also study the case of finite temperature and how the critical fields are changed by it. We will work in the quasi classical approximation always,valid for large volumes,in which the branes do not yet move significantly . We will make some comments on the issues involved  in the  last section.

The paper is organized as follows. In Section \ref{setup} we introduce the theory in the holographic setting, we derive the critical electric field first found in  \cite{Semenoff:2011ng}. In Section \ref{gyro} we discuss the same situation but with a background magnetic field instead.  We show that the gyromagnetic instability, which we would naively expect from weak coupling, is instead not there at all. In  Section \ref{pairproduction} we discuss pair productions for generic cases, electric field into $W$ bosons and magnetic field into monopoles, by computing the Euclidean bounce solution.  In Section \ref{mixed} we analyze the mixed problem in which both electric and magnetic fields are present simultaneously. In Section  \ref{thermal} we add temperature and study the thermal phase diagram.  We conclude in   Section \ref{conclusion} with some open questions.

\section{Holographic Setting}
\label{setup}

For a non-Abelian theory that contains charged particles and admits a holographic description we consider $\N=4$ SYM in the Coulomb branch with symmetry breaking $SU(N+1) \to SU(N) \times U(1)$. This is the simplest holographic realization of the Schwinger effect were the unbroken  $U(1)$ is the electro-magnetism and the massive W bosons are the charged particles to be pair created. The theory also admits monopoles and dyons and thus a whole set of generalized versions of the Schwinger effect. The holographic setup is valid in the limit of large $N$ and large 't Hooft coupling $\lambda = g^2 N$ where the $SU(N)$ unbroken part is replaced by its geometric dual given by type IIB string theory on $AdS_5 \times S^5$
\begin{align}
\label{coord}
ds^2 = L^2 \left( \frac{dr^2}{r^2} +r^2 \, dx^\mu dx_\mu +  d\Omega_5^2 \right)
\end{align}
where also $N$ units of Ramond-Ramond flux pass through the $S^5$ sphere.  The remnant $U(1)$ is described in the bulk by a physical D3-brane located at a certain radius of AdS $r_0$, later to be related to the vev $\langle\phi\rangle$ in the dual theory. Note that the coordinate $r$ has dimension of energy in these conventions. The gauge/gravity duality relates the parameters of the bulk string theory theory, AdS radius $L$ string coupling $g_s$ and string length $l_s$,  with the ones of the dual theory, gauge coupling $g$ or 't Hooft coupling $\lambda$ and $N$ by $L^2/l_s^2 = \sqrt{\lambda}/2\pi$ and   $g_s = g^2/4 \pi$.
Making a vacuum choice in the Coulomb branch is dual to putting a D3 brane in the bulk, at a certain radius $r_0$ and at a certain fixed point in the $S^5$ sphere.
The $W$-bosons correspond to fundamental strings F$1$ stretched between the isolated D3-brane  and the Poincar\'e horizon at $r=0$ with its mass given by the integrated tension 
\beq
\label{stringstreched}
\frac{1}{l_s^2} \int_0^{r_0} \sqrt{- {\rm det} \,  h_{ab}}= \frac{L^2 r_0}{l_s^2} = \frac{\sqrt{\lambda} r_0}{2 \pi} \ .
\eeq
where $h_{ab}={\rm diag}( - L^2r^2 , L^2/r^2)$ is the embedded worldsheet metric.
This has to be equal to the gauge theory mass $m_W= g v$ where $v$ is the expectation value of the adjoint field and provides the relation between the bulk and boundary variables $r_0$ and $v$ given by
\beq
v = \frac{ L^2 r_0}{2 l_s^2 \sqrt{\pi g_s }}
\eeq
The Higgs breaking corresponds to moving along the Coulomb branch at $\langle \phi \rangle = v t_{U(1)}$ where the generator is 
\beq
\label{generator}
t_{U(1)} = \frac{1}{\sqrt{2N(N+1)}}{\rm diag} (N, -1,\dots ,-1) \ .
\eeq 
The Coulomb branch is a flat direction in $\N=4$ SYM.  The gravitational force pulls the brane toward the infrared region $r=0$ but the RR flux, which jumps from $N$ to $N+1$ as we cross the brane at $r_0$, provides the balance repulsive force.  Thus $r_0$ is also a flat modulus in the bulk description. The position of the D3 brane in the $S^5$ sphere corresponds to some Higgs field, being in the vector representation of $SO(6)_R$, acquiring an expectation value. This could also be read from the fall off of the scalar field in the bulk dual to the Higgs field in the boundary.

Locally one can always approximate a curved metric as a flat space-time metric. For AdS we can do so by taking slices around a given radial position $r \pm \delta$ with a certain thickness $\delta$.   The metric is essentially Minkowsky flat provided $\delta$ is not too big.   Making the change of coordinate $r \to r' = L r /r_0$, $x_{\mu} \to x_{\mu}'=L  r_0  x_{\mu} $ we go into a frame where the metric is manifestly $\eta_{\mu\nu}$ and in these coordinates  $\delta' \ll L$ is the condition for local flatness; in the normal coordinates this is equivalently given by the condition  $\delta  \ll  r_0$. 
Local properties of the D3-brane can be understood just by zooming into this flat space-time slice,  this being   trustworthy as long as the string excitations which terminate on the brane do not wonder out of the strip $ r_0 \pm \delta$.  Here we can take the DBI action for the isolated D3-brane which represents the unbroken $U(1)$ is
\beq
\label{DBIaction}
S_{DBI} = \frac{1}{g_s l_s^4} \int d^4 x'\sqrt{-{\rm det}(\eta_{\mu\nu} - l_s^2 F_{\mu\nu, \ loc})} \ 
\eeq
where the suffix $loc$ stands for the local $\eta_{\mu\nu}$ frame. For a constant electric field the integrand reduces to $\sqrt{1-l_s^2 E_{loc}^2}$ and becomes imaginary above the critical electric field $E_{loc ,\ cr}=1/l_s^2 $  equal  to  the string tension.  Changing back  coordinates from $r',x_{\mu}'$ to  $r,x_{\mu}$  we have to properly rescale the electric field $ E \to E_{loc} = E / L^2 r_0^2$. This gives the critical electric field as measured from the original coordinates (\ref{coord}):
\beq
\label{Ecrdue}
E_{cr} = \frac{r_0^2 L^2}{l_s^2} = \frac{2 \pi m^2}{\sqrt{\lambda}} \ .
\eeq
This is interpreted in \cite{Semenoff:2011ng} as the critical field in the dual theory where pair production barrier drops to zero. Note that this derivation  is entirely local, just a rescaling  with the appropriate redshift factors from the local inertial frame to the original one. Moreover it is not about charged strings pair creation but neutral strings criticalities. Nevertheless we will see later that the global derivation leads to the same answer. The basic reason is that the Euclidean solution for the pair production is more and more localized near the brane as we reach the critical value.

\section{Absence of Gyromagnetic Instability}
\label{gyro}

Now we discuss the magnetic field background and the disappearance of the gyromagnetic instability, in the same setting as the previous section.  The spectrum of open string in constant $B$ background and flat space-time is solvable exactly \cite{Nesterenko:1989pz,Abouelsaood:1986gd,Ferrara:1993sq}. We will first review those results and then discuss them in AdS.

Let us discuss first the case of a bosonic open string with charges $q_1$ and $q_2$ at the two ends and $q=q_1+q_2$ the total charge of the string.
 \beq
S =  \int d\tau d\sigma {\cal L}- q_1 \int \left.d\tau A_{\mu} \partial_{\tau} X^{\mu}\right|_{\sigma=\sigma_1}  + q_2 \int \left.d\tau A_{\mu} \partial_{\tau} X^{\mu}\right|_{\sigma=\sigma_2}
\eeq 
where ${\cal L} = $ is the free string action and $\sigma_{i=1,2}$ refers to the two endpoints with charges $q_{i=1,2}$.  We add a background  magnetic field $F_{12} = B$. We can also consider a general case in which the ends  of the string are on two Dp branes at distance $d$ with Dirichlet boundary conditions for the coordinates $x_{p+1, \dots 26}$. The distance $d$ is related in the field theory to the value of the Higgs field vev $v= d/ 4 \pi^{3/2} \alpha' g_s$.  The string spectrum can be computed exactly since the presence of  $B$  does not affect the bulk equation of motion for the string but only the boundary conditions on the two ends.  The result for the bosonic spectrum is \cite{Abouelsaood:1986gd,Ferrara:1993sq}
\bea
\label{spectrumstring}
\alpha' {\cal E}^2 &=&  n  \sum_{n=1}^{\infty} ( a_n^{\dagger} a_n +  b_n^{\dagger} b_n) - \epsilon  \sum_{n=1}^{\infty} ( a_n^{\dagger} a_n - b_n^{\dagger} b_n) + \epsilon b_0^{\dagger} b_0  \nonumber \\
&& - 1 + \frac{1}{2}\epsilon(1-\epsilon)  + \frac{d^2}{4 \pi^2 \alpha'}
\eea
where $a_n$ and $b_n$ are the mode expansion in the coordinates affected by the magnetic field $x^1 \pm i x^2$ and have ordinary commutation relations. The dimensionless parameter $\epsilon$ is given  by 
\beq
\label{epsilonfactor}
\epsilon = \frac{1}{\pi} |\arctan{2 \pi \alpha' q_1 B} + \arctan{2 \pi \alpha' q_2 B}|
\eeq
and interpolates between $\epsilon \simeq 2 \alpha' q  B$ for $\alpha' q_i B \ll 1$ and   $\epsilon \to 1$  for   $\alpha' q_i B \to \infty$.  In the formula (\ref{spectrumstring}) we have omitted all the possible  excitations generated by transverse string fluctuations $\alpha^{\perp}_n$ orthogonal to the $B$ field  which are not changed with respect to the free string case and can be put to their ground state for simplicity.
The spin operator in the $12$-direction is
\beq
S =  \sum_{n=1}^{\infty} ( a_n^{\dagger} a_n -  b_n^{\dagger} b_n) \ ,
\eeq
and  the Landau level is $b_0^{\dagger} b_0 = N$. 
For states at a given spin value $S$ we are interested in the ones which have minimal energy.  Consequently we consider only the excited modes $b_0^{\dagger} b_0$ to be the Landau level and  $a_1^{\dagger} a_1 $ to be the spin and so  we can rewrite (\ref{spectrumstring}) as
\beq
\label{spectrums}
\alpha' {\cal E}^2 = \frac{1}{2}\epsilon(1-\epsilon) +\epsilon N  + (1-\epsilon) S + \frac{d^2}{4 \pi^2 \alpha'} - 1 \ .
\eeq
This corresponds to the spin-down choice, the spin-up would be instead obtained by exciting only $b_1^{\dagger} b_1$.
In the weak field limit $\epsilon \ll 1$ this reduces to 
\beq
\label{spectrumfieldfromstring}
{\cal E}^2 =  (2N +1) q B  - 2 q B S  + \frac{d^2}{4 \pi^2 \alpha'^2} + \frac{S - 1}{\alpha'}
\eeq
and this is precisely equivalent to (\ref{spectrumfield}) with gyromagnetic factor $g_s =2$ for every spin state.  Note that  states with only $a_1^\dagger a_1 \neq 0$  correspond to a minimal Regge trajectory whose  mass at zero magnetic field is  given by
\beq
M_S^2 =  \frac{d^2}{4 \pi^2 \alpha'^2} + \frac{S - 1 }{\alpha'} \ .
\eeq
The $W$ boson is the one with spin $S=1$, it is the first state in the Regge trajectory and becomes massless for the case of coincident branes at zero distance $d=0$. We are ignoring here  the zero spin state $S = 0$ which is a tachyon $B=0$, $d=0$. This state is the usual bosonic string tachyon and shall be projected out in the superstring setting.

We can now  discuss the tachyonic instability induced by the magnetic field.  In the field theory limit (\ref{spectrumfieldfromstring}), for any given spin state in the Regge trajectory  $S \geq 1$,  and any given distance $d$ between the two Dp branes,  there is always a critical B field at which this state becomes massless and above which it is tachyonic:
\beq
B_{cr}(S,d) = \frac{ d^2  +4 \pi^2 \alpha' ( S - 1  )}{4 \pi^2 \alpha'^2  q ( 2 S -1)} \ .
\eeq 
This $B_{cr}(S,d)$ is a monotonic function of $S$  and the lowest value, i.e. the first  criticality,  is  when  the spin $1$ state  becomes tachyonic
\beq
B_{cr}(d) = \frac{ d^2  }{4 \pi^2 \alpha'^2  q} \ .
\eeq 
This is  the gyromagnetic instability of (\ref{Bgyro}).  
However when $B_{cr}(d)$ is big enough, then the small field approximation $\epsilon \ll 1$ is no longer valid and the exact string formula for the spectrum (\ref{spectrumstring}) should be used instead of its field theory approximation (\ref{spectrumfieldfromstring}). The source of the tachyonic instability is in the Zeeman term $- 2 q B S$ in (\ref{spectrumfieldfromstring}) which comes from the term $-\epsilon S$ in the string formula (\ref{spectrumstring}). Since $\epsilon$ is saturating to a constant for large $B$ (\ref{epsilonfactor}), we expect a much milder instability in string theory than in field theory. 
To check if there are  criticalities, we may first  send $B \to \infty$, i.e.  $\epsilon \to 1$, and then compute  the distance $d_{cr}$ at which a criticality disappears.  This is given by $
d_{cr} = 2 \pi \sqrt{\alpha'}$.
This means that the gyromagnetic instability is completely absent once  the inter-brane distance becomes  bigger than a critical distance which is of the order of  string scale $l_s$. Above this distance, no matter how large the value of $B$ is  and whatever the value of $S$ is, there are no tachyons in the spectrum.

We can give a physical interpretation of this effect. When $\epsilon \ll 1$ the result (\ref{spectrumfieldfromstring}) is the same as the field theory in which the entire string fluctuating between the two branes corresponds to a particle with some mass, spin and gyromagnetic factor.  This can be understood comparing the time scales in the system. The string spectrum, without a  $B$ field,  is $M_S = d/2\pi \alpha'+ \dots$ where the dots contain all the possible excited oscillators of the free string.   The Larmor frequency for these massive states, considered as a definite  particle now,  in the $B$ field background is $\omega_{Larmor} = q B/2 M_S$.  This has to be compared to the frequency needed to see the {\it internal} structure of the string state which is that  of a generic fluctuation to propagate from one brane side to the other $\omega_{internal}=1/d$. When $\omega_{Larmor} \ll \omega_{internal}$ we can effectively consider  the whole string state  as a definite particle with a certain mass $M_S$ moving in the $B$ field background at a much lower frequency  than the one required to see its internal structure, and this is precisely the condition $q B \alpha' \ll 1$.

When $q B \alpha' \gg 1$ we are instead in a completely different regime.  $w_{Larmor}$ is much greater than $w_{internal}$ and consequently the string state
cannot be considered anymore as a free string moving slowly in the
magnetic field background.
The result of the exact computation (\ref{spectrumstring}) tells us that the contribution to the  mass squared coming from the Zeeman interaction ceases to grow with $B$ and instead saturates to the constant. 
We can describe in more detail how the string states enter the tachyonic instability from (\ref{spectrums}) \cite{Ferrara:1993sq}.  The slope of the Regge trajectory is set by the $(1-\epsilon)  / \alpha'$ and so it is always positive and becomes asymptotically flat in the limit  $B \to \infty$.  A state in a given Regge trajectory  becomes tachyonic once $\epsilon$ reaches the value
\beq
\epsilon_{cr}(N,S) = \frac{1}{2} +N -S + \sqrt{ \left( \frac{1}{2} + N  -S\right)^2 + 2 \left( \frac{d^2}{4 \pi^2 \alpha'^2} + S -1 \right)} \ .
\eeq
In order for this to correspond to a real value  $B_{cr}$ it has to satisfy the condition $\epsilon_{cr} <1$. This is possible to achieve only for the first Regge trajectory corresponding to the ground state in the Landau levels $N=0$. 
Moreover all the states in this fundamental trajectory have the chance to become tachyonic for a certain value of $B$.  The $\epsilon_{cr} (0,S)$ is growing with $S$ and reaching $1$ asymptotically
\beq
\epsilon_{cr}(0,1) = -\frac{1}{2} + \sqrt{\frac{1}{4} + \frac{d^2}{2 \pi^2 \alpha'^2} }\ , \qquad \dots  \qquad \epsilon_{cr}(0,\infty) = 1 \ .
\eeq
So  we have established that, for distance lower than the critical $d < d_{cr}$, all the states in the first Regge trajectory, and only those, can  become tachyonic for increasing values of $B$. The slope of the trajectories flattens as $B \to \infty$.  For $d > d_{cr}$ none of the string states become tachyonic, for whatever value of the magnetic field.\footnote{$4d$ closed string background may also have exotic behavior under application of a constant magnetic field, see for example in Heterotic string theory \cite{Kiritsis:1995iu}.}

\begin{figure}[h!t]
\epsfxsize=10.5cm
\centerline{\epsfbox{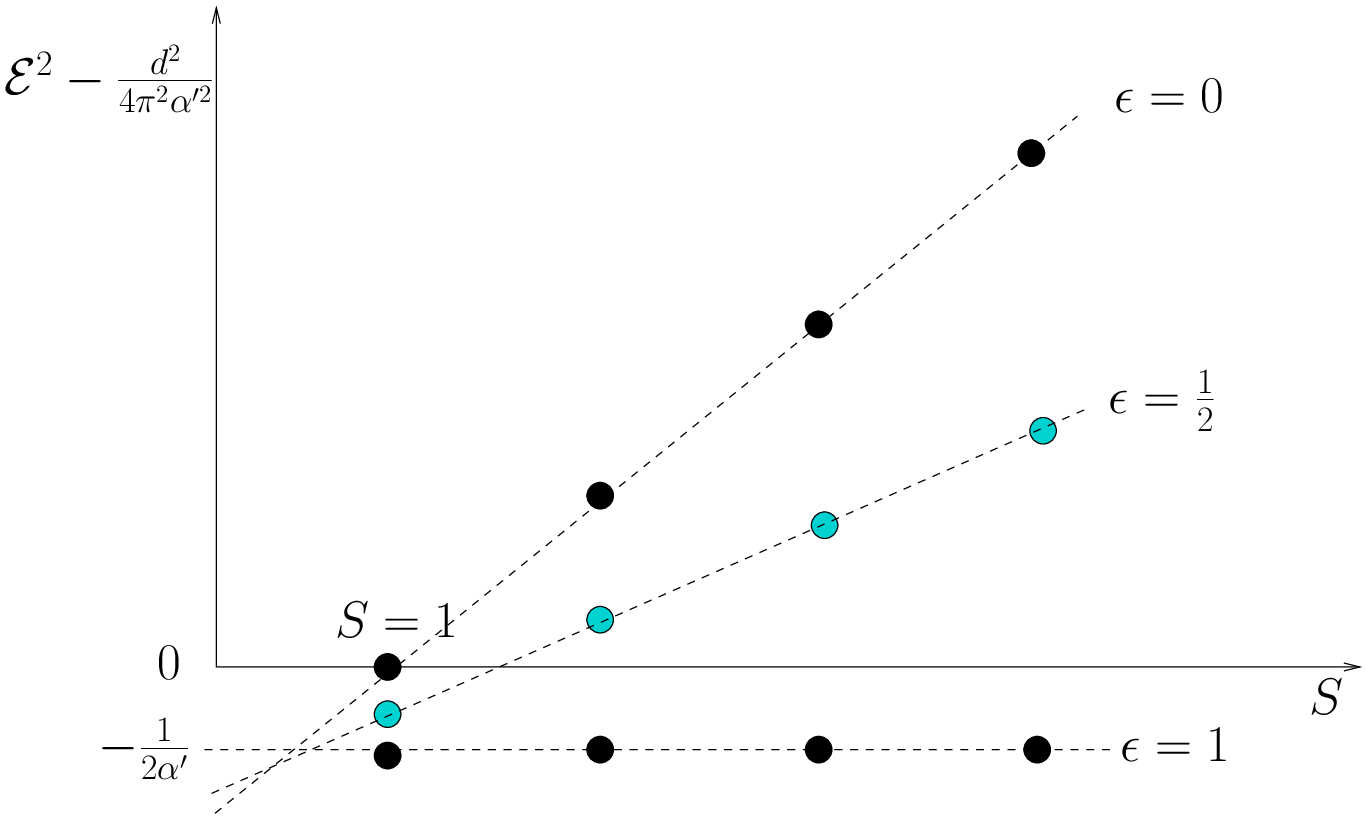}}
\caption{{\footnotesize       Evolution of the first Regge trajectory, spin-down, spectrum as $B$ goes from $0$ to $\infty$. The first state $S=1$ is the $W$ boson. The trajectory slope flattens and becomes asymptotically flat as $\epsilon \to 1$. The negative mass squared contribution becomes asymptotically  constant.   }}
\label{firstregge}
\end{figure}

For superstrings there is little difference. 
In the Ramond sector there are no magnetic instabilities at all. 
In the Neveu-Schwartz sector, where there are the W bosons,  the field theory instability is recovered for small values of $B$. 
The energy for the first Regge trajectory is  
\beq
\label{massspectrumsupersymm}
\alpha' {\cal E}^2 = -\frac{\epsilon}{2}  + \epsilon N +  (1-\epsilon) (S - 1) + \frac{d^2}{4 \pi^2 \alpha'}  \ .
\eeq
which is slightly different from the bosonic counter part (\ref{spectrums}), but has all the same qualitative features: the first trajectory $N=0$ is the only one that can become tachyonic, trajectories are flat as $B \to \infty$, and most important above a critical distance $d_{cr} = \pi \sqrt{2 \alpha'}$ no tachyons are allowed for any value of $B$ (see Figure \ref{firstregge}).

Electric and magnetic instabilities can be also understood via T-duality.  In a T-dual perspective the electric field is a relative tilt between the two branes in space-time and critically arises when the tilt is equal to the speed of light.
In the case of the magnetic field the T-dual correspond to a tilt of an angle $\theta$ in space directions where the angle is related to the magnetic field by $\theta =\pi \epsilon$. There is indeed an instability \cite{Epple:2003xt,Berkooz:1996km,Hashimoto:1997gm}  related to the  relative tilting.  Note that the tilt cannot exceed $\theta=\pi$ which is equivalent to a brane  parallel to an  anti-brane and this is the  geometric counterpart of the saturation limit $\epsilon \to 1$ as $B \to \infty$. Using $\theta = \pi$ as an upper bound on the amount of negative contribution due to the tilt, we find a critical distance $d_{cr} = \pi  \sqrt{2 \alpha'} $ above which there are no tachyons.

Now we return to our original problem. We have  a stack of  $N$ D3 branes and one isolated D3 separated by a distance $d$. A $B$ field proportional to generator (\ref{generator}) is turned on and the F1 strings stretched between the two branes have charges which in the large $N$ limit can be taken to be  $q_1 = 1/\sqrt{2} + {\cal O}(1/N) $ and $ q_2 = {\cal O}(1/N)$, where $q_1$ and $q_2$ refers respectively to the isolated D3 and the stack of $N$  D3's.   We are interested in the spectrum of this configuration in a curved  AdS geometry  (\ref{coord}) where with the $N$ branes are at the far infrared  $r =0$ and the D3 at $r=r_0$.  We can consider an intermediate situation  in which the $N$ D3 branes are placed at a generic radius $r_*$ between the infrared and $r_0$ with $0 \leq r_* \leq r_0$. This generic configuration interpolates between the problem in  which we are primarily interested in , which is the limit $r_* =0$ and a situation in which the distance between the branes is so small that we can neglect the curvature of AdS. This is when $r_0 -r_* \ll r_0$  and in this range we can use the previously given solution in flat space-time for the string spectrum.   Here we know that the gyromagnetic instability is there only at very short distance and disappears when the distance, in the $\eta_{\mu\nu}$ frame, reaches the critical value. This is equivalent in the original coordinates to   
\beq
\label{criticaldistancer}
(r_0 -r_{*})_{cr}  \simeq  \frac{r_0 l_s}{2 \sqrt{\pi} L} \ .  
\eeq
First note  that  $r_0 -r_{*, \ cr} \ll r_0$, and this very important fact means that  one is still inside the safe zone for the flat space-time approximation to be valid.   Second, and most importantly,  the physical situation in which we are interested, $r_* =0$ is well above the critical distance and so much above any tachyonic instability. 
By continuity, we expect that even as $r_{*} \to 0$ we still remain outside the critical distance.\footnote{Note that this argument poses no restriction on $r_0$, it just has to be greater than zero.}
Note that there is a small caveat in this argument. The region of interest $r_* \to 0$ is well outside the flat space time safe zone. To complete the argument we need some extra information, and this is that the ground state energy is
\beq
\label{stringstrecheddue}
\frac{1}{l_s^2} \int_{r_*}^{r_0} \sqrt{- {\rm det} \,  h_{ab}}= \frac{\sqrt{\lambda} (r_0-r_*)}{2 \pi} \ .
\eeq
 Given this, and the fact that the critical distance is inside the flat space-time region and can be computed (\ref{criticaldistancer}), we can argue that $r_* = 0$ does not have any tachyonic instability, whatever the value of $B$.

We express the maximal shift of energy due to the magnetic field in terms of field theory parameters. 
The energy squared spectrum of W bosons, in the $B \to \infty$ limit, is at most modified by a term which is at most $\simeq 1/\alpha'$ in the local $\eta_{\mu\nu}$ frame. This follows from (\ref{massspectrumsupersymm}) in the $\epsilon \to 1$ limit which becomes $ {\cal E}^2 = -1/2 \alpha' + d^2/4 \pi^2 \alpha'^2$. This formula holds in the flat space-time case. Embedding in AdS may add a order one coefficient in front of the negative
term, but will not effect its non leading character. Expressing this in terms of dual boundary variables we get
\beq
{\cal E}_W^2  =  m_W^2  \left( 1 -  {\cal O}\left(\frac{1}{\lambda^{1/4}}\right) \right)
\eeq
The disappearance of the tachyonic instability is thus a large $\lambda$ effect in the field theory description as can be seem by the fact that the critical distance (\ref{criticaldistancer}) is smaller with respect to $r_0$ by a factor $\propto 1/\lambda^{1/4}$.

\section{Pair Production}
\label{pairproduction}

In this section we compute the rate of pair production of electro magnetically charged particles in constant electric and magnetic fields. 
We review the results in flat space time and adapt them to the case of theories which have an AdS dual.
The calculation for a constant magnetic field is new. 

The method requires finding solutions of a Wick rotated theory which when rotated back to Minkowski space correspond to
a $q$-$\bar{q}$ pair created at sufficient distance to escape the barrier. We will do it first for the field theory and then for the holographic case.

We will adopt a general approach such to apply both to W and monopoles. We consider a particle with mass $m$ ($m_W$ and $m_M$ eventually), charged under a field $F$ ($E$ or $B$), with a coupling $q$ ($g$ or $\widetilde{g}$). 
This method of computing the particle pair creation rate has been used in \cite{Affleck:1981bma,Affleck:1981ag}; we will look for a ``worldline bounce''.
The action for the worldline particle is
\bea
\label{worldlineaction}
S_{E} &=& m \int d\tau \sqrt{{X'}^2} - q \int A_{\mu} dX^{\mu} \nonumber \\
     &=& m {\cal P} - q F {\cal A }_{\parallel} 
\eea
where $F = d A$ and in the second line  we assumed the trajectory to be closed with a certain perimeter ${\cal P}$ and an area ${\cal A}_{\parallel}$ along the field $F_{ij}$.
The bounce is a loop in the Euclidean space which  extremises the action. 
Given a generic loop we can always minimize the action by projecting the entire orbit to the $F_{ij}$ plane; this will shorten the perimeter ${\cal P}$ by keeping fixed ${\cal A}_{\parallel}$.  We can then minimize further $S_{E}$ by taking the shape with maximal area for a fixed  perimeter, which a circular shape with radius $R$. The circular loop so obtained depends only one variable $R$ and the action becomes
\beq
\label{fieldtheory}
S = m  2 \pi R   - q F  \pi R^2
\eeq
This time the extremum is not a minimum but a maximum.  The extremum is at the classical solution
\beq
\label{smallB}
R_{cl} = \frac{m}{q F} \qquad S_{cl}= \frac{\pi m^2}{q F}
\eeq
this extremised solution has one negative eigenvalue of the quadratic action when expanded around it. It is the signal of an instability when interpreted as a tunneling in the Minkowski space-time. A part from the translational zero modes, all the other eigenvalues are positive. 
In the Euclidean formulation the trajectories of point like particles in a constant background are closed circles in a plane perpendicular to $F$ as opposed to the constantly accelerated hyperboloids in Minkowski. 
This corresponds to the pair produced particles which, once produced above the barrier,  recede to each other with a constant acceleration. The pair production probability is given by
\beq
\label{fieldmainformula}
w \propto e^{- S_{cl}}
\eeq
with the pre-exponent factor given by the determinant of the positive modes.

We can then apply the formula to the two specific examples: W-bosons and  monopoles pair production. Formula (\ref{fieldmainformula}) gives the probability of pair production (\ref{pairpw}) and (\ref{pairpm}) respectively. In both cases we have to take care that the worldline approximation is valid, and that is that the bounce radius $R_{cl}$ must be bigger than the particle size. For the W boson we have to take the Compton wave length for the size, and $R_{ct} \simeq 1/m_W$ exactly when $E$ becomes of order of $E_{Schw}$. For the monopole instead we have to use its classical radius $r_M \simeq 1/M_W$  which is $1/g^2$ bigger than the Compton length. The worldline approximation breaks down at $ B \ll g v^2$ which is much smaller than $B_{Schw}$ and is by the way of the same order of the perturbative instability $B_{gyro}$. So as long as the vacuum is stable the worldline approximation is a good one.

Now let us analyze the problem in the holographic dual side.  The $W$ boson is replaced in the bulk by an F1 string whose boundaries are located at the D3 brane and the Poincare horizon.  For the monopole we just have to replace the F1 with the D1 string, and the Wilson with the 't Hooft loop.  Dyons are given by bound states F1-D1.
Geometrically the problem is the same for all cases and again can be treated in a unified way (see for previous computation of this kind \cite{Gorsky:2001up}).
We will take a string with tension $T$ ($1/l_s^2$ for the F1 or $1/g_s l_s^2$ for the D1). The Euclidean configuration is a string worldsheet with one circular  boundary at the D3 brane at radius $r_0$, the ``worldsheet bounce''. The boundary of the worldsheet is a loop with charge $q$ under the field $F$ on the brane worldvolume. The Euclidean action is
\beq
S_{E} = \T \int d\sigma d\tau \sqrt{{\rm det} \ g_{2}(\sigma,\tau)}  - q \int_{boundary} dX^{\mu} A_{\mu}
\eeq
where the first part is the Nambu-Goto action with $g_{2}(\sigma,\tau)$ the pull-back metric and the second couples the boundary charge to the field $F = dA$. This is a generalized version of the field theory action (\ref{worldlineaction}) and we will see that in the weak field limit it gives the same result.

\begin{figure}[h!t]
\epsfxsize=5.5cm
\centerline{\epsfbox{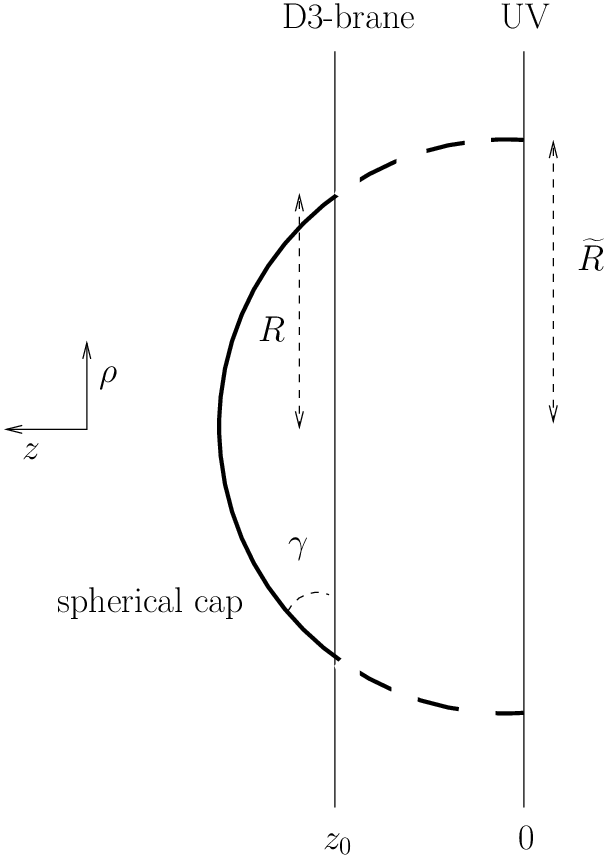}}
\caption{{\footnotesize Geometry of the worldsheet bounce. The dashed part is not physical, just a continuation of a minimal surface solution in AdS to the UV boundary. }}
\label{geometry}
\end{figure}

We can use radial coordinates $\rho,\theta$ in the plane $F_{ij}$ and assume that the solution will be invariant and only function of $\rho$.   
The geometry of the solution is a circular cap surface with a radial profile given  in Figure \ref{geometry}. It is a surface which extremise the area and ends on a loop of radius $R$ on the D3-brane. As before we first minimize the Euclidean action in all infinite directions apart from the size $R$ which at the end must be maximized. 
For this problem it is convenient to move to the coordinates $z=1/r$ where the metric is manifestly conformally flat
\beq
ds^2 = \frac{L^2}{z^2} \left( dz^2 + dx_{\mu}dx^{\mu} \right)
\eeq
The Euclidean action is then
\beq
\label{actiond1string}
S_{E} = \T \int_0^{R} d\rho 2\pi \rho  \left(\frac{L}{z(\rho)}\right)^2  \sqrt{1+ z'(\rho)^2 } -q F \pi R^2
\eeq
where the profile $z(\rho)$ is the one to be determined. 
A  minimal surface in hyperbolic space is  given by a half sphere 
\beq
\label{circlesolution}
z(\rho) = \sqrt{ \widetilde{R}^2 - \rho^2 }
\eeq
These are the stationary solutions to the first part of the action (\ref{actiond1string}).
 This curve should be truncated at $z=z_0$ where the string ends on the D3 brane, since the part from $0$ to $z_0$ is not physical and $z_0 = z(R)$.
 
The curve is fixed once we determine the integration constant, \ie the radius $\widetilde{R}$ of the sphere. The radius $R$ is measured at $z=z_0$ and is given by $R^2 + z_0^2 =\widetilde{R}^2 $. 
The action as a function of $R$ has the following expression:
\beq
\label{actionER}
S_E=\T 2\pi L^2 \left( \sqrt{1+ \frac{R^2}{z_0^2}}-1 \right) -q F \pi R^2
\eeq
The maximization with respect to $R$ is equivalent to  the  following balance of forces
\beq
\T \cos(\gamma) = q  F_{loc} 
\eeq
The radius at the stationary point is given by
\beq 
R_{cl} = z_0  \sqrt{\left(\frac{\T L^2 }{q F z_0^2}\right)^2 - 1} \ ,
\eeq
and it leads to the action
\beq
S_{E \, cl} = q F \pi z_0^2 \left(\left(\frac{\T L^2 }{q F z_0^2}\right)   - 1 \right)^2 \ .
\eeq
A critical point is reached when radius $R$ and classical action vanish. This happens at the following critical value for the field 
\beq
\label{criticalF}
q F_{cr} = \frac{\T L^2}{ z_0^2}
\eeq
where the radius $R$ and also the classical action vanishes. This is when the sphere in Fig. \ref{geometry} becomes exactly tangent to the D3 brane and nothing is left for the physical cap. The easiest way to see what is the reason of the existence of this criticality is to look again at the action as function of $R$ (\ref{actionER}). At large $R$ is always dominated by the electro-magnetic coupling and thus negative $-q F \pi R^2$.  At small $R$ we can expand it and we have 
\beq
S_E =  \frac{\T \pi L^2 R^2}{z_0^2} -q F \pi R^2  + \dots
\eeq
this is positive as long as $F < F_{cr}$ and so we have a barrier. When $F = F_{cr}$ the barrier disappears and the vacuum becomes unstable.  It is the stringy  nature that makes the small $R$ behavior to be proportional to $ \T R^2$ and thus in direct competition with the electro-magnetic coupling. For particles instead (\ref{fieldtheory}) the small $R$ behavior is proportional to $ m R $ and thus  there is always a barrier. Quantum correction to the Euclidean bounce have been computed in \cite{Ambjorn:2011wz} where they showed a the existence of a sub-leading correction in $\lambda$ but no qualitative change.\footnote{In this respect it would be nice to see if the same quantum correction can be recovered by studying higher loops terms in the DBI action.}

The weak field limit $F \ll F_{cr}$ correspond to the field theory limit. In this case the action becomes (\ref{fieldtheory}) with the mass
\beq
m = \frac{\T L^2}{z_0}
\eeq
which is like (\ref{stringstreched}) the one of a string with tension $T$ stretched from $z_0$ to the horizon at $z \to \infty$.

The result can be applied both to the W pair production and to the monopole pair production. For the W we have $\T=1/l_s^2$ and $q=1$ from which we recover exactly the critical field predicted by the DBI action (\ref{Ecrdue}). We now see why a purely local quantity gives the correct answer, the classical solution is in fact localized near the D3 brane as $E \to E_{cr}$. For the magnetic case we have to use the D1 tension which is $1/g_s \gg 1/\alpha'$ greater than the F1 tension. But the charge $q$ is different also and is now equal to $4\pi/g^2 =1/g_s$. These two factor cancel exactly  in (\ref{criticalF}) and this gives a critical magnetic field exactly equal to the critical electric field $B_{cr} = E_{cr}$. The point is that the D1 is heavier but also the coupling to the $B$ field is larger. The breaking point is at the same threshold (see also  \cite{Gorsky:2001up}).

Some comments related to S-duality are this context is in order. S-duality being a non perturbative symmetry is generically broken if one maintains only the leading term in large N large λ expansion, in particular the S-dual theory is not expected to be at weak coupling.  The leading terms in the pair-production probability turn out to be  S-dual and this to a certain extent is a surprise, as already noticed in [7]. We have not found at this stage a BPS like argument to explain this.
On the other hand note that one has not found any reason to expect that any of these systems has a perturbative
instability thus an expectation that in an S Dual theory there is no magnetic perturbative instability while also not rigorous , could be well entertained.

For the electric case we have two independent approaches which give the same answer. Thus we can be very confident that the $E_{cr}$ is indeed a physical critical point. For the magnetic field we do not see any sign of instability from the DBI action. The DBI action integrand is $\sqrt{1 + l_s^2 B_{loc}^2 }$ for a constant magnetic field as opposed to $ \sqrt{1 - l_s^2 E_{loc}^2 }$ for the electric case. We thus do not see any local $B_{cr}$ above which the action would  cease to make sense. But the DBI action on a flat brane is not required to know a D1-strings pair production. Monopoles are in fact known to be well described near the D3 branes also as BIons, which are  spikes of the D3 brane and are solution of the non-linear DBI equation \cite{Gibbons:1997xz,Callan:1997kz}. The Euclidean bounce involving a monopole loop would thus correspond to a D3 brane with non-trivial topology.
Note also the according to \cite{Tseytlin:1996it,Green:1996qg} 
 the S-dual of the DBI action of the D3-brane is self-dual, i.e. it  has the same functional form,  and thus would predict the same critical value for the field strength.

A qualitative argument for the emergence of the $E_{cr}$ has been advocated in \cite{Semenoff:2011ng} also.  The disappearance of the tunneling barrier can be understood as a consequence of the electro-magnetic potential between the two particles $q-\bar{q}$.  The potential for a $W-\bar{W}$ pair created  at a distance $d$ is $
V_{eff \ W}(d) = 2 m_W -  E  d - \alpha/d $ where the last term is the attractive potential and the coefficient $\alpha$ can be taken from the W-boson Wilson loop on the boundary of AdS and it is $\alpha = 4\pi^2 \sqrt{\lambda}/\Gamma^4(1/4)$. The barrier disappears when $V_{eff\ W}(d) = V_{eff\ W}'(d)=0$ which happens for $d_{cr} = \alpha/m_W$ and  $E_{cr} = m^2/\alpha \simeq .7 \times 2 \pi m_W^2/\sqrt{\lambda} $. This is a crude approximation to the Euclidean bounce, for which we have instead  $d_{cr}=0$ and $E_{cr} = 2 \pi m_W^2/\sqrt{\lambda} $, but it gives nevertheless a possible intuitive interpretation from the dual boundary theory perspective.\footnote{The very same argument could also be applied for weakly coupled theories, such as QED. It would predict also for this a critical field, but incredibly big  $E_{cr} \simeq E_{Schw}/ g^2$. }  The very same argument is applicable also for  monopoles pair production.  In this case the potential is  $
V_{eff \ M}(d) = 2 m_M -   B d 4\pi / g^2 -  \alpha/ g_s d $  where  the extra $g_s$ in the final term is necessary  to convert the Wilson loop into a 't Hooft loop.    So one obtain, up to an over all multiplicative factor, that  $V_{eff \ M} = V_{eff \ W} \,  4\pi /g^2$ and so $B_{cr} = E_{cr}$.

\section{Mixed $E$ and $B$}
\label{mixed}

We can now ask what is the effect of a combination of magnetic and electric fields on the pair production rate and the related instability.  There are two effects which should be  taken  into account.  The first one is an indirect effect of the magnetic field on the pair production of W-bosons which is made manifest by relativistic invariance. One can decompose the magnetic field with respect to the direction of the electric field direction, a parallel component $B_{\parallel}$ along the electric field and a
perpendicular one $B_{\perp}$. The DBI determinant inside the square root of the action (\ref{DBIaction}) in the local frame is
\beq
\label{detdbi}
- {\rm det} \left( \begin{array}{cccc}-1&l_s^2 E_{loc}&&\\-l_s^2E_{loc}&1&l_s^2 B_{loc \ \perp}&\\&-l_s^2B_{loc \ \perp}&1&l_s^2B_{loc \ \parallel}\\&&-l_s^2B_{loc \ \parallel}&1 \end{array}\right)
\eeq 
which is 
\beq
1-l_s^4 (E^2_{loc} - B_{loc \ \parallel}^2-B_{loc \ \perp}^2) + l_s^8 E_{loc}^2 B_{loc \ \parallel}^2 \ .
\eeq
The second term is just the invariant $F_{\mu\nu}F^{\mu\nu}$, the third is a higher order term $F^4$. 
The critical field value is that which makes (\ref{detdbi}) vanish
\beq
E_{loc \ cr} = \frac{1}{l_s^2} \sqrt{\frac{1+l_s^4 ( B_{loc \ \parallel}^2 + B_{loc \ \perp}^2)}{1+l_s^4 B_{loc \ \parallel}^2}} \ .
\eeq
Bringing it back to the original frame and expressing it in terms of the boundary field theory parameters leads to 
\beq
\label{ebcrite}
E_{cr} = \frac{2 \pi m^2}{\sqrt{\lambda}} \sqrt{1+ \frac {B_{\perp}^2}{\frac{4 \pi^2 m^4}{\lambda}+ B_{\parallel}^2}} \ .
\eeq
We see that the   critical value of $E$  is in general increased  by the presence of a $B$ field.
There is no change at all if the perpendicular component vanishes, as it can be seen by the fact that (\ref{detdbi}) factorizes once that component vanishes. The parallel component has also an higher order effect in mitigating  the enhancement due to the perpendicular one. 
The previous result can also be obtained using Lorentz invariance, first by boosting to a reference frame where $B_{\perp} = 0$ and $E_{cr}$ is $1/l_s^2$ independently of $B_{\parallel}$, and then boosting back to the original frame. The square root in (\ref{ebcrite}) is a Lorentz transformation factor. This problem was also considered in \cite{Acatrinei:2000qm} for branes in flat space-time.

\begin{figure}[h!t]
\epsfxsize=6.0cm
\centerline{\epsfbox{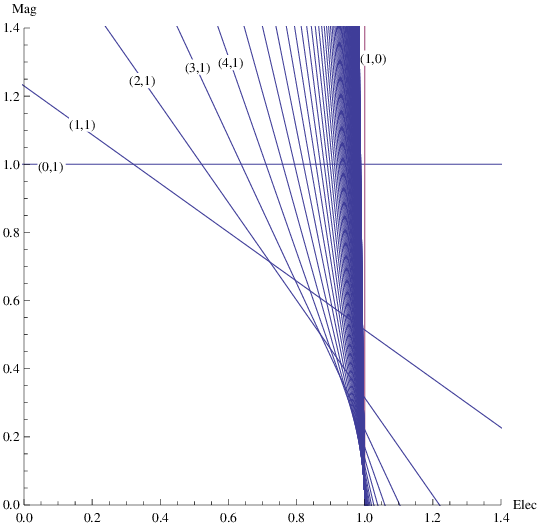}}
\caption{{\footnotesize  Sub-critical zone in the $E$-$B$ plane in the frame where they are parallel to each other. The various lines correspond to Eq. (\ref{dyonsformula}) for all the states in the spectrum.}}
\label{mixedEB}
\end{figure}

A second effect to be considered is that with generic $E$ and $B$ fields any kind of dyonic state can also be pair produced. To understand the $E$-$B$ phase diagram we need to take into account all of them. 
The first thing to do is to get rid of the previously discussed effect and go to a boosted frame where $E$ and $B$ are parallel. We have then reduced the problem  to a two dimensional phase diagram. Then we have to take into account the dyonic states in the spectrum.  Let us denote a dyon with charges $N = (n_e,n_m)$ so that the  W bosons is a $(1,0)$ and the monopole a $(0,1)$. The mass is given by  
\beq
\label{dyonmass}
m_{(n_e,n_m)} = v g \sqrt{n_e^2  + \frac{16 \pi^2 n_m^2}{g^4}} \ .
\eeq
Finally we compute the pair production probability for any generic $E$-$B$ and any generic $(n_e,n_m)$. The functional to extremise is always of the form (\ref{actiond1string}). The tension is given by the bound state of D1-F1 strings
\beq
\T_{(n_e,n_m)}= \frac{1}{l_s^2} \sqrt{n_e^2  + \frac{n_m^2}{g_s^2}}
\eeq
which is proportional to the particle mass (\ref{dyonmass}). The last piece to be defined is the force on the extremities which was denoted as $q F$ in the action (\ref{actiond1string}) and now is given by the sum of the two forces being them parallel to each other
 \beq
q F  =  g E + \frac{4 \pi B}{g} \ .
\eeq
Then we finally use the critical field computed in (\ref{criticalF}) and we have
\beq
\label{dyonsformula}
E n_e + B \frac{4 \pi n_m}{g^2} = \frac{2 \pi  m^2}{\sqrt{\lambda}}  \sqrt{ n_e^2 + \left(\frac{4\pi n_m}{g^2} \right)^2 }
\eeq
Note that  this is manifestly $SL(2,Z)$ invariant. This equation should be intended as defining a critical line in the $E$-$B$ plane due to the dyon $(n_e,n_m)$.  Actually in the semiclassical spectrum we have only the states with charges $(n,1)$ and $(1,0)$.  Combining all the particle in the spectrum we have the phase diagram in Figure \ref{mixedEB}. The region of the phase diagram which is safe from any criticalities is the one contained below the envelop of the various lines. Figure \ref{mixedEB} correspond to a particular choice of the coupling $g=2$. The shape of the no-critical region is coupling dependent, in particular it interpolates between the circle and the square as the coupling $g$ goes from zero to infinity:
\bea
\sqrt{E^2 + B^2} \leq  \frac{2 \pi  m^2}{\sqrt{\lambda}}  &\qquad& g \to 0 \nonumber \\ 
|E|,|B| \leq \frac{2 \pi  m^2}{\sqrt{\lambda}}   &\qquad&   g \to \infty \ . 
\eea

\section{Thermal Excursion}
\label{thermal}

In the presence of temperature the critical behavior is changed, we  study the modifications of the  pair production and criticality features  at finite temperature.
A thermal state corresponds in the bulk to the Schwarzshild AdS black hole
\beq
\label{bhmetric}
ds^2 = L^2 \left(-\left(r^2-\frac{r_h^4}{r^2}\right) \, dt^2  + \frac{dr^2}{\left(r^2 - \frac{r_h^4}{r^2}\right)} + r^2  dx^i dx_i +  d\Omega_5^2 \right)
\eeq
with horizon at $r_h$ and temperature of the dual boundary theory given  by
\beq
\label{bht}
T = \frac{r_h}{\pi}
\eeq
The probe D3-brane sits  at $r_0$ and the  mass of the W boson is still given by Eq. (\ref{stringstreched}), see also \cite{Tseytlin:1998cq,Kiritsis:1999tx} for earlier  works on the same configuration.  At finite temperature there is no more cancellation of forces and the probe brane feels a net attractive force toward the black-hole. Here we will not have to deal with this effect since we shall be mostly interested in finding the critical value of $E$ and $B$ fields. 

\begin{figure}[h!t]
\centerline{
\epsfxsize=6.0cm
\epsfbox{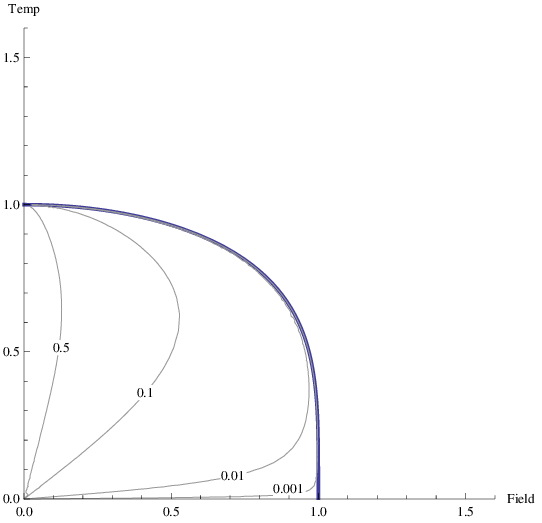}
\qquad
\epsfxsize=6.0cm
\epsfbox{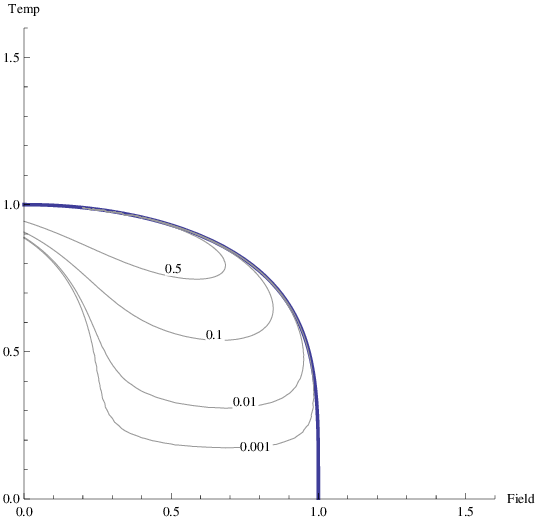}}
\caption{{\footnotesize Phase diagram $F$-$T$. Contour plots respectively of  the parameter expansion (\ref{cond1}) for the left panel and (\ref{cond2}) for the right panel. }}
\label{ETphase}
\end{figure}

We first derive the critical electric field as in Section~\ref{setup}, which means that we will compute it locally with the opportune scaling factors. 
The effect of the presence of an electric field effect is to shift the string tension. The effective string tension is locally given by:
\begin{align}
\label{Teff}
\T_{eff,\ loc} =  \frac{1}{\ell_s^2} \left(1 - \ell_s^4 E_{loc}^2 \right)
\end{align}
with 
\beq
\label{Elocgrav}
E_{loc} = - \frac{E}{\sqrt{-g_{00} g_{ii}}} =  \frac{E}{L^2 r_0^2 \sqrt{1 - \frac{r_h^4}{r_0^4}}} 
\eeq
the dependence on the temperature enters in $g_{00}$.
The critical electric field is thus given by
\bea
\label{criticalEwithT}
E_{cr}(T) &=& \frac{L^2 r_0^2}{l_s^2} \sqrt{1-\frac{r_h^4}{r_0^4}} \nonumber \\
&=& \frac{2 \pi m^2}{\sqrt{\lambda}} \sqrt{1-\frac{T^4 \lambda^2}{16 m^4}}
\eea
expressed in both bulk and boundary quantities.
This gives a curve in the $(E,T)$ plane that interpolates between the zero temperature critical $E_{cr}$ of  (\ref{Ecrdue}) and the temperature at which the horizon coincide with the D3-brane position $r_h = r_0$. 
The sub-critical zone is the one inside this curve (see Figure \ref{ETphase}).

Now we have to find the bounce solution like in Section \ref{pairproduction}. Again we expect the previous result to be correct because near the criticality the bounce solution should shrink to zero size and thus should be dependent only on local space-time properties. This will not only confirm the previous result for the electric field, but also give the temperature dependence of the critical magnetic field $B_{cr}(T)$.  By changing the coordinates by the transformation $z=1/r$ and continuing  to Euclidean space (\ref{bhmetric}) becomes
\beq
\label{bhmetricze}
ds^2 = \frac{L^2}{z^2} \left(\left(1-\frac{z^4}{z_h^4}\right) \, d\tau^2  + \frac{dz^2}{\left(1-\frac{z^4}{z_h^4}\right)} +  dx^i dx^i  \right)
\eeq
The Euclidean time is compactifyed with $
\tau = \tau + \pi z_h$ in order for the geometry to be smooth at the horizon leading to  thermal behavior (\ref{bht}).
We search for the classical  stationary solution of the string action (\ref{actiond1string}) in this new metric.
The bounce is not longer circular symmetrical in $x,\tau$, so that a generic solution would require to solve a partial differential equation.  But if we want just to  study the near to critical regime, i.e. to check the critical line (\ref{criticalEwithT}), the shape of the bounce is roughly ellipsoidal, and can be made circular with a change of coordinates.  We call the factor  
\beq
\gamma(z) = 1-\frac{z^4}{z_h^4} \ ,
\eeq
and we do the following change of coordinates
\beq
\label{gammachange}
\tau_c =  \gamma(z)   \tau \qquad \qquad  x_c = \sqrt{\gamma(z)} x 
\eeq
Then the metric becomes
\beq
\label{bhmetriczegamma'}
ds^2 = \frac{L^2}{z^2 \gamma(z)} \left(d\tau_c^2  + \left(1 + \frac{\gamma'(z)^2 \tau_c^2}{\gamma(z)^2} + \frac{\gamma'(z)^2 x_c^2}{4 \gamma(z)} \right) dz^2   +  dx_c^2   \right)
\eeq
where $\gamma'(z)$ is the derivative respect to $z$.
Neglecting the terms with  derivative of $\gamma$, later to be checked when possible,  the metric simplifies to 
\beq
ds^2 = \frac{L^2}{z^2 \gamma(z)} \left( d\tau_c^2  + dz^2 +  dx_c^2   +  \dots \right)  
\eeq
We can solve the bounce by using a circular symmetric ansatz with a profile $z(\rho)$ given by a slightly deformed version of (\ref{actiond1string}):
\beq
\label{actiond1stringtemp}
S_{E} = \T \int_0^{R} d\rho 2\pi \rho  \frac{L^2}{z(\rho)^2 \left(1-\frac{z(\rho)^4}{z_h^4}\right) }  \sqrt{1+ z'(\rho)^2 } - q F_c \pi R^2
\eeq
From the change of variables in (\ref{gammachange}) 
\beq
F_c = \frac{F}{\gamma(z)^{3/2}} \ .
\eeq
Since the bounce is circular $\tau_c = x_c =R$ at most. And the two conditions for $\gamma'$ terms to be negligible are satisfied by the most stringent one
\beq
\label{condgamma'}
R \ll \frac{\gamma(z)}{\gamma'(z)} = \frac{z_h^4}{4 z^3}  \left(1-\frac{z^4}{z_h^4}\right)  \ .
\eeq
Another condition to impose is that the circle of the bounce $R$ is smaller than the compactification scale of $\tau$. This condition is 
\beq
\label{conditionr}
R \ll   \frac{\pi z_h}{2}  \left(1-\frac{z^4}{z_h^4}\right) 
\eeq
and this is a  stronger inequality  than (\ref{condgamma'}).

Yet we can make a further approximation. Let us consider the terms $L^2 / z(\rho)^2$ in the functional action (\ref{actiond1string}), this is modified into  $L^2  /z(\rho)^2 \gamma(z(\rho))$ in (\ref{actiond1stringtemp}) and this makes no longer valid the nice integrable solution (\ref{circlesolution}). But when the bounce $z$-thickness $\delta z = z_{max} -z_0$ is not too deep, to be quantified later, we can reduce exactly to the functional (\ref{actiond1string}).
Making the following change of coordinates
\beq
\tz = z - \frac{2 z_0^5}{z_0^4 + z_h^4} \qquad \qquad  \tL = \frac{\tz_0}{z_0 \sqrt{\gamma(z_0)}}\  L 
\eeq
which is just a translation in $z$ and a rescaling of the AdS radius $L$, the canonical functional
\beq
\label{actiond1stringtempchanged}
S_{E} = \T \int_0^{R} d\rho 2\pi \rho   \frac{\tL^2}{\tz(\rho)^2}  \sqrt{1+ \tilde{z}'(\rho)^2 } - q  F_c \pi R^2
\eeq
has the same local behavior of (\ref{actiond1stringtemp}) provided
\beq
1- \frac{\tL^2}{\tz^2} \frac{z^2 \gamma(z)}{L^2}  \ll 1 
\eeq
which after some rearrangements reduces to the condition
\beq
\label{conditionz}
1-\frac{\gamma(z)^3 }{ \gamma(z_0) \left(1+\frac{z_0^4}{z_h^4} - \frac{2 z_0^5}{z z_h^4}\right)^2} \ll 1
\eeq
We can then  use the solution of Section \ref{pairproduction} and we have for the bounce radius
\bea
\label{radiusfinalt}
R_{cl} &=& \tz_0  \sqrt{\left(\frac{ \T \tL^2 }{q F_c \tz_0^2}\right)^2 - 1} \\ 
&=& \frac{z_0 (-z_0^4 + z_h^4)}{z_0^4 + z_h^4} \sqrt{\left(\frac{\T L^2 }{\gamma(z_0) q F_c z_0^2}\right)^2 - 1}
\eea
and for the critical field  
\beq
\label{criticalEwithTverifyed}
F_{cr,\  c} = \frac{ \T L^2}{q \gamma(z_0)  z_0^2} \qquad \Rightarrow  \qquad  F_{cr} = \frac{\T L^2 \sqrt{ \gamma(z_0)}}{q  z_0^2}
\eeq
which then confirms the local derivation (\ref{criticalEwithT}).

We can check that the two approximations become in fact increasingly good near the critical line. The first condition is that the bounce radius is smaller than the compactification radius of the Euclidean time $\tau$ (\ref{conditionr}). This is also strong enough to imply the (\ref{condgamma'}) condition regarding the smallness of the $\gamma'$ terms in the metric (\ref{bhmetriczegamma'}). Using (\ref{radiusfinalt}) we can rewrite (\ref{conditionr}) as
\beq
\label{cond1}
\frac{2 T \sqrt{\lambda} m}{\pi \left(1+ \frac{T^4 \lambda^2}{16 m^4} \right)} \sqrt{\frac{F_{cr}(T)^2}{F^2} -1} \ll 1
\eeq
which is increasingly well satisfied as $F \to F_{cr}(T)$. The other condition (\ref{conditionz}) should be evaluated at the tip of the bounce given by $\tz_{max} = \tz_0 F_{cr} / F$, and finally it becomes
\beq
\label{cond2}
1- \frac{\left( 1-  \frac{T^4 \lambda^2 F^4}{16 m^4 F_{cr}(T)^4} \right)^3}{ \left( 1- \frac{T^4 \lambda^2}{16 m^4}\right)\left( 1+ \frac{T^4 \lambda^2}{16 m^4}-2 \frac{T^4 \lambda^2 F }{16 m^4 F_{cr}(T)}  \right)^2} \ll 1
\eeq
which again is increasingly well satisfied as $F \to F_{cr}(T)$.
So we have an entire region, inside the sub-critical zone and close to the critical line $E_{cr}(T)$  which can be treated in this approximation. In Figure \ref{ETphase} this is represented with normalization $F_{cr}(T=0)=1$ and $T_{cr}=1$. We also plot the lines where the conditions (\ref{cond1}) and (\ref{cond2}) become of order one and thus are violated.

The formula (\ref{criticalEwithTverifyed}) can then be applied to any configuration in which dyons are pair produced by parallel $E$ and $B$ and formula (\ref{dyonsformula}) corrected by thermal effects becomes  
\beq
\label{dyonsformuladue}
E n_e + B \frac{4 \pi m_m}{g^2} = \frac{2 \pi  m^2}{\sqrt{\lambda}} \sqrt{1-\frac{T^4 \lambda^2}{16 m^4}}  \sqrt{ n_e^2 + \left(\frac{4\pi n_m}{g^2} \right)^2 }
\eeq
There is thus a universal correction proportional to $\sqrt{\gamma(r_0)} =  \sqrt{1-\frac{T^4 \lambda^2}{16 m^4}}$ and the domain of sub-critical $E$ and $B$ has the same shape of Figure \ref{mixedEB} and is opportunely rescaled by $\sqrt{\gamma(r_0)}$.

\section{Conclusions and Open Questions}
\label{conclusion}

We discussed some issues related to Schwinger pair creation, electric and magnetic, in a context where the holographic description is weakly coupled. In the electric case it was shown in \cite{Semenoff:2011ng} that a critical electric field exists at which the pair production barrier drops to zero. In the magnetic analog we showed that again the stringy nature brings about a surprise.  A critical instability which is generically present at weak coupling in field theory, manifested by  the emergence of a tachyonic ground state in the W boson spectrum at the value $B_{gyro}$ of the magnetic field, disappears completely in the holographic set up. The trend is opposite to that of the electric field but the underlying reason is the same: the $\alpha'$ corrections.  We have deployed the same method of \cite{Gorsky:2001up,Semenoff:2011ng} to study in a unified way the pair creation for W bosons and for monopoles or any other dyonic state in the spectrum. In the last part of the paper we described the generalized phase diagram with $E$, $B$ and also their temperature dependence.

There are a number of issues which require a better understanding.
One involves the argument for the absence of the gyromagnetic instability. For that
we relied on the exactly known spectrum of open strings in the background
magnetic field in flat space,  in particular we showed that in the AdS case the
effective system of branes are separated by a distance which is well above the distance threshold above which the tachyons disappear.
The argument is rather convincing but does not constitute
 a  rigorous proof. It would be interesting to uncover the  exact spectrum of open strings in an
AdS and magnetic field background on the D3 brane, perhaps by using integrability properties,  to
confirm this statement. Moreover it would be interesting to study also the
non-Abelian case in which, at weak coupling, the analog of the gyromagnetic
instability is known to occur due to a tachyon in the charged gluons spectrum \cite{Nielsen:1978rm}. The argument of Section \ref{gyro} is applicable only to the Coulomb phase, but it seems to suggest that the gyromagnetic instability should be absent also in the non-Abelian phase.

Another open issue regards the critical field $B_{cr}$ found in Sec. \ref{pairproduction}. The unified analysis of pair creation suggests that there should be a critical magnetic field when the D1 string gets broken. Unfortunately in this regime, the treatment of Sec. \ref{pairproduction} becomes questionable. Indeed, since the bounce action evaluates to zero, a semiclassical expansion is no longer obviously valid.  For the electric case one has a backup argument since  also the DBI action has an instability which gives the same result.  For the magnetic case the DBI action shows no signs of an instability at $B_{cr}$. This may be due to the fact that in the analysis in  Sec. \ref{pairproduction} we  neglected the back-reaction of the string on the D3 brane on which it ends, and this may be significant for the D1 string.   To compute the pair creation probability near $B_{cr}$ we should thus deploy a more powerful method which does take into account the back-reaction.


Finally we want to discuss the problem of the ultimate fate of the brane when  electric or magnetic fields are turned on. We said that the BPS-ness of the system is broken by the field, and the gravitational force is  no longer balanced by the RR flux force. This implies a net force toward the IR region of AdS. In other words a brane with a field turned on is rolling toward the $r \to 0$, just by classical forces with a certain time scale.
On the other hand we have the Schwinger pair production effect.
 First of all we want to stress that the computation of the pair production probability, which has been done considering the brane as static, always makes sense for appropriately large volumes.  Note that the classical dynamics has a natural time scale for the brane to reach the IR. The pair production is instead a probability per unit of time and per unit of volume.  For sufficiently large volumes, pairs are pair produced much before the brane starts to move.  So pair production can always be considered as a quasi-static process. On the other hand an important question to ask is how the pair production is affecting the classical fall of the brane. This is likely to be more and more important near the critical value for the fields $F_{cr}$ where the barrier for the pair production drops and the production becomes classical, and no longer suppressed by quantum tunneling. To determine the ultimate fate of the brane is a harder problem left for the future, and involves understanding of the back reaction of the pair produced particles on the brane, and in particular the dynamics of pair production near the critical value when the barrier drops.

\section*{Acknowledgments}

We thank for discussions C.~Bachas, F.~Bruckmann, J.~P.~Gauntlett, D.~Israel, Z.~Komargodski, K.~Lee, N.~Seiberg and A.~Tseytlin.
This work is partially supported by the American-Israeli Bi-National Science Foundation and the Israel Science Foundation Center of Excellence.  SB is founded also by the Lady Davies fellowship of the Hebrew University. ER thanks also  the Leverhulme foundation and the theoretical physics group at Imperial college for its hospitality.

\end{document}